\documentclass[sigconf]{acmart}

\setcopyright{none}
\renewcommand\footnotetextcopyrightpermission[1]{}

\AtBeginDocument{%
  }

\setcopyright{acmcopyright}
\copyrightyear{2024}
\acmYear{2024}
\acmDOI{XXXXXXX.XXXXXXX}

\acmConference[CIDR '24]{Conference on Innovative Data Systems Research}{January 14--17, 2024}{Chaminade, USA}
\acmPrice{15.00}
\acmISBN{978-1-4503-XXXX-X/18/06}

\usepackage{caption}
\usepackage{subcaption}
\usepackage{listings}
\usepackage{wrapfig}
\usepackage{enumitem}

\newcommand{\smalltt}[1]{{\texttt{\small #1}}}

\newcommand{\ind}[1]{VMTree}

\newcommand{\veh}[1]{Shortcut-EH}

\usepackage{tabularx}
\newcolumntype{L}[1]{>{\raggedright\let\newline\\\arraybackslash\hspace{0pt}}m{#1}}
\newcolumntype{C}[1]{>{\centering\let\newline\\\arraybackslash\hspace{0pt}}m{#1}}
\newcolumntype{R}[1]{>{\raggedleft\let\newline\\\arraybackslash\hspace{0pt}}m{#1}}

\definecolor{darkgreen}{RGB}{28,171,70}
\definecolor{darkred}{RGB}{182,21,14}
\definecolor{darkyellow}{RGB}{255,165,0}

\newcommand{\red}[1]{\noindent{\color{red} #1}}
\newcommand{\green}[1]{\noindent{\color{darkgreen} #1}}
\newcommand{\yellow}[1]{\noindent{\color{darkyellow} #1}}

\begin{document}
\title{Taking the Shortcut: Actively Incorporating the Virtual Memory Index of the OS to Hardware-Accelerate Database Indexing}

\author{Felix Schuhknecht}
\affiliation{%
  \institution{Johannes Gutenberg University Mainz}
  \city{}
  \country{}
}
\email{schuhknecht@uni-mainz.de}

\begin{abstract}

Index structures often materialize one or multiple levels of explicit indirections (aka pointers) to allow for a quick traversal to the data of interest. Unfortunately, dereferencing a pointer to go from one level to the other is costly since additionally to following the address, it involves two address translations from virtual memory to physical memory under the hood. In the worst case, such an address translation is resolved by an index access itself, namely by a lookup into the page table, a central hardware-accelerated index structure of the OS.
However, if the page table is anyways constantly queried, it raises the question whether we can \textit{actively incorporate} it into our database indexes and make it work for us. Precisely, instead of materializing indirections in form of pointers, we propose to express these indirections directly in the page table wherever possible. By introducing such \textit{shortcuts}, we (a)~effectively reduce the height of traversal during lookups and (b)~exploit the hardware-acceleration of lookups in the page table.  
In this work, we analyze the strengths and considerations of this approach and showcase its effectiveness at the case of the real-world indexing scheme extendible hashing.

\end{abstract}

\settopmatter{printfolios=true}

\maketitle
\pagestyle{plain}

\vspace*{-0.2cm}
\textbf{Citing this research:} \\ This version is a preprint of the paper accepted for publication at CIDR 2024. When citing this research, please cite the CIDR version.\\

\noindent \textbf{Code available:} \\ Code and artifacts of this project are available under: \\ \url{https://gitlab.rlp.net/fschuhkn/taking-the-shortcut}

\section{Introduction}

\begin{figure*}[h!]
  \centering
  \begin{subfigure}[b]{.43\linewidth}
    \includegraphics[page=1, width=\linewidth, trim={0 3.6cm 61cm 0}, clip]{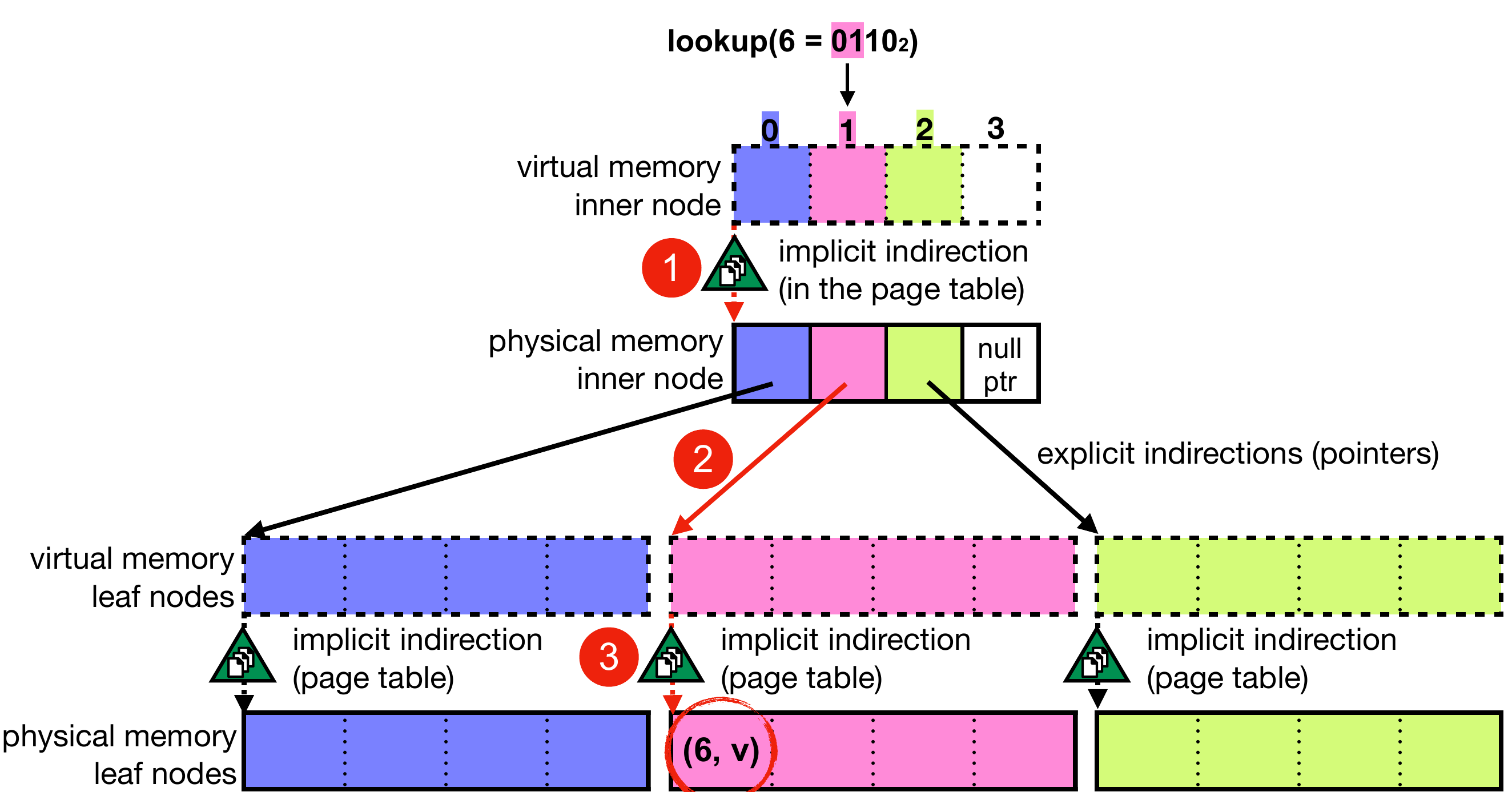}
    \caption{Traditional inner node: three indirections must be resolved.}
    \label{fig:basic_idea1}
  \end{subfigure}
  \hfill
  \begin{subfigure}[b]{.55\linewidth}
    \includegraphics[page=2, width=\linewidth, trim={0 4cm 49.5cm 0}, clip]{figures/general_indirections.pdf}
    \caption{Shortcut inner node: only a single indirection must be resolved.}
    \label{fig:basic_idea2}
  \end{subfigure}
  \vspace*{-0.3cm}
  \caption{Indexing three leaf nodes with a traditional pointer-based radix inner node (Figure~\ref{fig:basic_idea1}) versus a shortcut (Figure~\ref{fig:basic_idea2}).}
  \vspace*{-0.3cm}
  \label{fig:basic_idea}
\end{figure*}

Tree-based index structures implement some sort of hierarchy of inner nodes, which contain indirections that point to a portion of the next level. These indirections are often materialized explicitly as pointers containing virtual memory addresses. 
Unfortunately, dereferencing these pointers is surprisingly costly, as more is going on than what meets the eye. To visualize the problem, in Figure~\ref{fig:basic_idea1}, we depict the memory perspective for an exemplary radix-style inner node with four slots, where the first three slots contain pointers to leaf nodes on the next level.  
We can see that by default, both the inner node as well as all leaf nodes are represented by virtual \textit{and} physical memory. The reason for this is that when we allocate memory for our nodes (using \smalltt{malloc} or \smalltt{new}), we actually allocate only virtual memory. This virtual memory is transparently mapped to physical memory on page granularity, whereas this mapping is materialized in the page table, the index of the memory subsystem of the OS. 
As a consequence, our exemplary data structure already contains two levels of implicit indirections right after allocation. Additionally, by materializing pointers to the individual leaf nodes, we introduce a level of explicit indirections. 
This means that looking up a key such as 6 (\smalltt{0110} in binary) already requires the resolving of \textit{three} indirections in total: 
One implicit indirection when accessing the inner node, one explicit indirection when following the pointer, and another implicit indirection when accessing the leaf. 

\subsection{Taking the Shortcut}
\label{ssec:taking_the_shortcut}

This leads to the question whether we can create some sort of \textit{shortcut} to reduce the total number of indirections. In the end, we only want to map slots of an inner node to nodes on the next level. To express this, \textit{one} level of indirection should be sufficient. 

To achieve this, we propose to express the mapping from inner node slots to nodes on the next level purely using implicit indirections in the page table.  
Figure~\ref{fig:basic_idea2} shows the equivalent state to Figure~\ref{fig:basic_idea1} following this approach. The key difference is that instead of materializing both inner node and leaf nodes via virtual \textit{and} physical memory, we realize the inner node solely by virtual memory and the leaf nodes solely by physical memory. Then, instead of mapping inner node slots to leaf nodes via pointers, we map portions of the virtual memory representing the leaf node slots directly to the corresponding physical memory of the leaf nodes using a technique called memory rewiring~\cite{rewiring}. 
In total, establishing this shortcut effectively eliminates two levels of indirections in comparison with the traditional approach.
Even better, this resolving is not only performed automatically by the OS, it is also hardware-accelerated by the CPU. 

\begin{figure}[h!]
\vspace*{-0.2cm}
\includegraphics[width=.9\columnwidth]{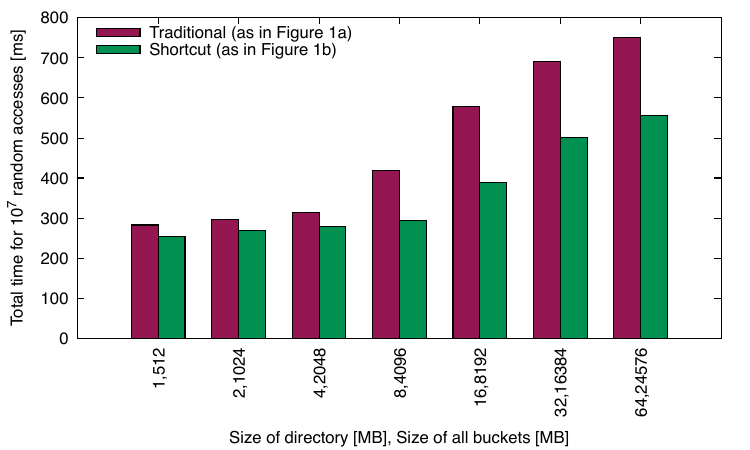}
\vspace*{-0.2cm}
\caption{Effect of taking the shortcut under $10^7$ uniformly distributed random accesses.}
\vspace*{-0.3cm}
\label{fig:random_access}
\end{figure}

To get an impression of how much we can gain, we implemented a traditional radix-style inner node/leaf node relationship (Figure~\ref{fig:basic_idea1}) and a corresponding shortcut variant (Figure~\ref{fig:basic_idea2}). We then compare their performance under a sequence of lookups while varying the number of indexed leaf nodes of size~$4$KB (the size of a small memory page). 
Figure~\ref{fig:random_access} shows the results. As we can see, eliminating indirections via a shortcut has a significant positive impact on the lookup performance. Also, we can see that the positive effect depends on the total size of all inner nodes: The higher the fan-out, the more the random accesses going through the traditional variant become a bottleneck.
Index structures that could benefit from shortcuts are those that (a)~use page size nodes and (b)~perform a radix-style traversal. An application that meets both criteria is for instance the well-known extendible hashing~\cite{eh, multi_level_eh}, which utilizes a directory to adaptively index hash buckets. Thus, we will integrate and showcase shortcuts in extendible hashing in Section~\ref{sec:eh} and show how to significantly reduce the access overhead of its directory during lookups.

\section{Shortcuts in Practice}
\label{sec:shortcuts}

Before diving into how to create and use shortcuts, let us discuss the page table and how to manipulate the memory mappings expressed therein.
The page table sits at the core of virtual memory management and is essentially a coarse-granular radix tree, which maps virtual memory to physical memory at the granularity of pages. 
It offers two very interesting properties we want to exploit: 
First, modern CPUs implement hardware-accelerated lookups in the page table. Second, the CPU provides a hardware cache for address translations, called the \textit{Translation Lookaside Buffer (TLB)}, which caches the most recent address translations. By incorporating shortcuts, we automatically exploit both features in our indexes.

Interestingly, the memory mappings expressed in the page table can be actively manipulated at runtime from user space via a technique called memory rewiring~\cite{rewiring}. Precisely, it is possible to create new mappings from virtual to physical memory as well as to update existing mappings at page granularity. 
The core principle works as follows: In contrast to the traditional situation, where the programmer gets purely in contact with virtual memory, rewiring introduces handles to \textit{both} virtual and physical memory, where physical memory is realized as so-called \textit{main-memory files}. A main-memory file acts like a normal file, despite that it is backed by volatile (physical) main memory instead of disk pages. Thus, a main-memory file effectively provides a handle to physical memory. 
Using \smalltt{mmap} it is possible to create a virtual memory area that is mapped to such a memory-memory file. By this, we establish a controllable mapping from virtual to physical memory. 
This principle has been exploited successfully in the past to accelerate data structures~\cite{rewiring, pma, rewired_cracking}, snapshotting~\cite{rewiring, anyolap}, as well as table-scans~\cite{storageviews2, storageviews3}.

\subsection{Construction and Maintenance}

With the background knowledge at hand, let us now see how to practically create a new shortcut. 
For the following example, we assume there exists a traditional inner node with four slots referencing three leaf nodes via pointers, as shown in Figure~\ref{fig:basic_idea1}. We want to construct an equivalent shortcut as visualized in Figure~\ref{fig:basic_idea2}. 

Let us discuss the physical memory perspective first. 
To introduce mappable physical memory to user space, we maintain a self-managed pool of physical pages, which we represent by a single main-memory file. This main-memory file can resize on demand at page granularity to provide a flexible amount of physical memory to our application. 
To create and resize such a main-memory file, we utilize the following two system calls:

\begin{figure*}
    \centering
    \includegraphics[page=3, width=.95\linewidth, trim={0 0.8cm 15cm 0}, clip]{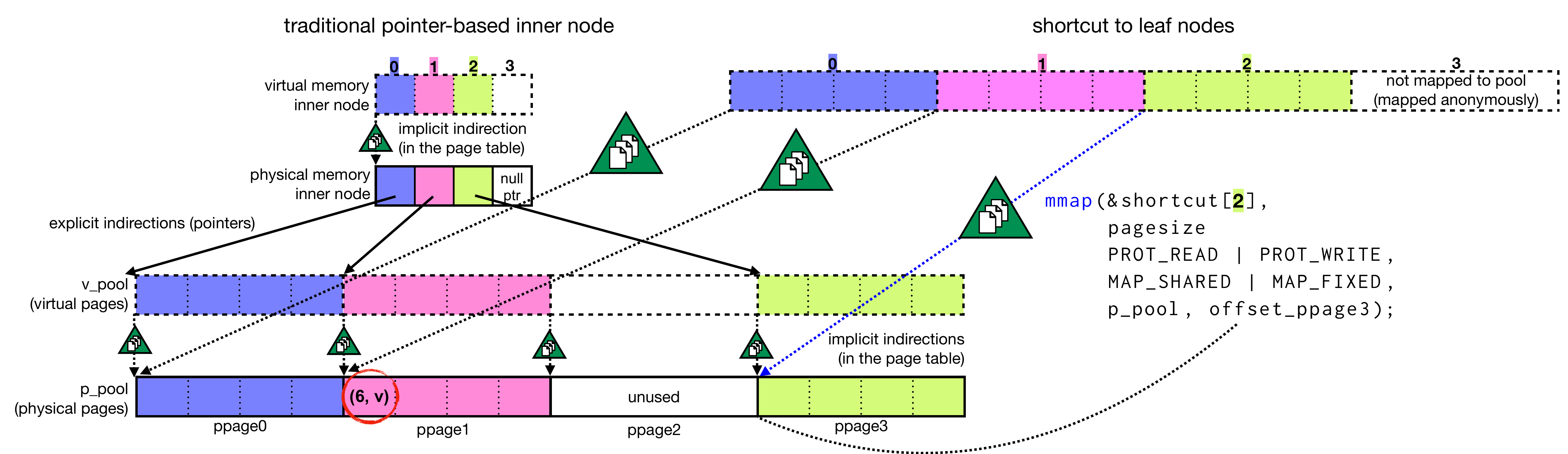}
            \vspace*{-0.2cm}
    \caption{Creating an equivalent shortcut inner node alongside a traditional inner node. To enable shortcuts, the physical memory of the nodes to create a shortcut to must be allocated from the page pool.}
        \vspace*{-0.3cm}
    \label{fig:shortcutting}
\end{figure*}

\noindent
\begin{minipage}{\linewidth}
\begin{footnotesize}
\begin{lstlisting}
int p_pool = memfd_create("pool", 0);      // create pool
ftruncate(p_pool, 4 * pagesize);     // resize to 4 pages
\end{lstlisting}
\end{footnotesize}
\end{minipage}

\noindent The call to \smalltt{memfd\_create()} creates the main-memory file and returns a so-called file descriptor (\smalltt{p\_pool}), which serves as our handle to the physical page pool in the following. 
To acquire more physical pages, we call \smalltt{ftruncate()} with the desired amount and initialize the new pages to avoid expensive hard page faults~\cite{rewiring} at access time later on. Note that a physical page can also become unused. If the unused page marks the end of the main-memory file and the pool size is above a specified threshold, we simply shrink the file using \smalltt{ftruncate()}. For all unused pages that are not located at the end of the main-memory file, we maintain a queue of offsets into the main-memory file to locate them quickly for reuse.
Additionally, at all times, we maintain a virtual memory area starting at address \smalltt{v\_pool} that maps linearly to the entire main-memory file, rendering it easily accessible. 
Note that to enable the creation of a shortcut referencing a set of leaf nodes, all physical memory of these leaves must originate from this page pool, as shown in  Figure~\ref{fig:shortcutting}. In the shown example, we assume that \smalltt{ppage0}, \smalltt{ppage1}, and \smalltt{ppage3} are the physical pages from the pool that represent our leaf nodes, while \smalltt{ppage2} is currently unused. 

To create a shortcut from an \smalltt{k}-slot inner node to its leaf nodes, as shown on the right side of Figure~\ref{fig:shortcutting}, we have to perform two steps: 
(1)~Reserve a consecutive virtual memory area of size \smalltt{k * pagesize} representing the shortcut, where each virtual page represents a slot.
(2)~Replicate the indirections of the traditional inner node by mapping each virtual page of the shortcut to the physical page of the corresponding leaf node. 
To perform step~(1), we call \smalltt{mmap()} and instruct it to create a virtual memory area that is backed by anonymous physical memory by passing the flags \smalltt{MAP\_PRIVATE} and \smalltt{MAP\_ANON}. This performs a mere reservation of a virtual memory area of \smalltt{k}~pages.
To replicate the indirection of the \smalltt{i}-th slot of the traditional inner node in step~(2) in our shortcut, we first have to identify the physical page of the pointed-to leaf node. To do so, we exploit that (a)~the physical page originates from \smalltt{p\_pool} and (b)~there exists a linear mapping between \smalltt{v\_pool} and \smalltt{p\_pool}. Therefore, we first retrieve the virtual page~\smalltt{v\_leaf} of the leaf node from the pointer in slot
~\smalltt{i}. Then, we compute $\smalltt{offset\_leaf} = \smalltt{v\_leaf}-\smalltt{v\_pool}$ to get the offset of the virtual page in \smalltt{v\_pool}. Due to the linear mapping, this offset also marks the beginning of the physical page of the leaf in \smalltt{p\_pool}.
Consequently, we can now map the \smalltt{i}-th virtual page of the shortcut representing slot~\smalltt{i} to the physical page of the leaf node at $\smalltt{offset\_leaf}$ using:

\noindent
\begin{minipage}{\linewidth}
\begin{footnotesize}
\begin{lstlisting}
mmap(&shortcut[i],      // update mapping at &shortcut[i]
     pagesize,                            // map one page
     PROT_READ | PROT_WRITE,               // permissions
     MAP_SHARED | MAP_FIXED,   // update existing mapping
     p_pool, offset_leaf);  // map to pool at offset_leaf
\end{lstlisting}
\end{footnotesize}
\end{minipage}
\noindent By passing \smalltt{\&shortcut[i]} as the first argument along with the flags \smalltt{MAP\_SHARED} and \smalltt{MAP\_FIXED}, we remap the corresponding virtual page, which is currently mapped anonymously, to the physical page specified by the passed offset \smalltt{offset\_leaf} into \smalltt{p\_pool}.
By executing the previously described remapping procedure for every slot, we iteratively build up our shortcut until all indirections are set. If we are in the lucky situation to map neighboring virtual pages of the shortcut to neighboring physical pages in the pool, we can do so in a single \smalltt{mmap()} call to minimize call overhead.
Note that to reflect updates, we simply execute step~(2) for each updated slot. 

\vspace*{-0.2cm}
\section{Considerations}
\label{sec:micro_benchmarks}
\vspace*{-0.1cm}

Before integrating shortcuts into an actual database index, we want to discuss how shortcuts behave in certain relevant situations. This (a)~will allow us to effectively use the technique in Section~\ref{sec:eh} and (b)~will be useful for other works that plan to incorporate shortcuts. 

We perform all of the following experiments on an Intel Core i7~12700KF with 32GB of DDR5-4800 RAM. The CPU has a hybrid design with 8 performance cores (with hyper threading) and 4 efficiency cores, where we turn off the efficiency cores for the experiments. The L1 TLB can cache 256 address translations for 4KB pages, whereas the L2 TLB can cache 3072 translations. The operating system is a vanilla 64-bit Ubuntu 22.04.

\subsection{Factor in the Cost of Creation!}
\label{ssec:beware1}

We start by analyzing the cost of creating a shortcut and subsequently using it in comparison with the traditional variant.
To do so, we set up a benchmark that measures the following steps: (1)~Allocate a new inner node with $n$~slots. 
(2)~Set $n$~indirections to $n$~individual leaf nodes. 
(3)~Optionally, eagerly populate the page table for the shortcut variant (by default, a page table entry is created lazily under the first access).  
(4)~Perform $10$M~random accesses on randomly selected leaf nodes located through the inner node.
(5) Perform the $10$M accesses of~(4) for a second time.  

We allocate a single inner node with $n=2^{22}$~slots in this experiment that resembles the situation of a wide inner node as we will face it in our application later on, or a large number of narrower inner nodes occurring on the same level.  
Table~\ref{tab:mmap} shows the results for the traditional variant and the shortcut variant (both lazy and eager population). We show normalized times here, i.e., the time to allocate, to set the indirections, and to populate is reported for a single page, whereas the access time is reported for a single access.  

\begin{table}[h!]
\vspace*{-0.2cm}
\small
    \centering
    \begin{tabular}{R{1.72cm}||C{1.45cm}|C{1.85cm}|C{2.02cm}}
                        & Traditional & Shortcut (lazy) & Shortcut (eager) \\ \hline\hline
Allocate [$\mu s$]      & \green{0.0}      & \green{0.0}          & \green{0.0} \\ \hline
Set Indir. [$\mu s$]    & \green{2.1}      & \red{447.5}    & \red{449.4} \\ \hline
Populate [$\mu s$]      & -           & -               & \yellow{74.1} \\ \hline
1. Access [$\mu s$]     & \yellow{22.6}      & \yellow{50.4}          & \green{16.5} \\ \hline
2. Access [$\mu s$]     & \yellow{23.0}      & \green{18.6}          & \green{16.7} \\
    \end{tabular}
    \caption{The normalized cost of creating and subsequently randomly accessing an inner node with $2^{22}$~slots.}
    \label{tab:mmap}
    \vspace*{-0.5cm}
\end{table}

As we can see, the allocation phase~(1) is basically for free, as it is a mere reservation of virtual memory.
Phase~(2), where we set the indirections to the leaf nodes, performs drastically different between the variants. While setting the pointers in the traditional variant causes negligible cost of only $2.1\mu s$, calling \smalltt{mmap()} to map a virtual page to its physical counterpart takes on average around $450\mu s$. 
This shows the price we have to pay for being able to take shortcuts: The initialization time is two orders of magnitude more costly, which is something we will have to factor in when incorporating shortcuts later on.
Next, phase~(3) optionally eagerly populates the page table and must be analyzed together with phase~(4) where we perform the accesses. As we can see, populating the page table before performing the accesses has the advantage that the first access becomes cheaper by a factor of $3$x.
Finally, in phase~(5), we can see that after the first round of accesses has happened, the second round of access performs almost equally, independent of whether the page table has been populated eagerly or lazily. 

Both observations indicate that the cost of handling shortcuts should be hidden if possible. In Section~\ref{sec:eh}, we implement this by creating, maintaining, and eagerly populating all shortcuts \textit{asynchronously} with respect to all modifications to the traditional index.

\vspace*{-0.2cm}
\subsection{Avoid TLB-Thrashing!}
\label{ssec:beware2}
\vspace*{-0.1cm}

Next, let us analyze the impact of the fan-in, i.e., the number of slots that index the \textit{same} leaf node. As we will face this specific situation in extendible hashing (and potentially in other index structures), we want to analyze performance implications in advance. 

To do so, we again allocate a wide inner node with $n=2^{22}$~slots. However, this time, we vary the total number of leaf slots such that multiple (neighboring) slots of the inner node refer to the same leaf node. After creation, we perform $10$M~random lookups through the structure and measure the total time.
\begin{figure}[h!]
\vspace*{-0.2cm}
\includegraphics[width=.89\linewidth, trim={0 0 0.6cm 0}, clip]{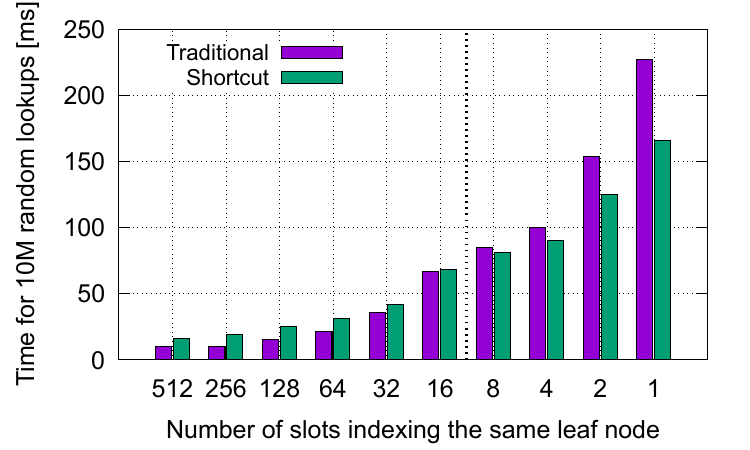}
\vspace*{-0.3cm}
\caption{Impact of fan-in.}
\label{fig:tlb_thrashing}
\vspace*{-0.3cm}
\end{figure}
Figure~\ref{fig:tlb_thrashing} shows the results for the traditional variant and the shortcut variant when varying the fan-in from $512$ (indexing $2^{13}$~leaf nodes) to $1$ (indexing $2^{22}$~leaf nodes). First of all, we observe that for both variants the runtime increases with the total amount of leaf nodes to index. While this is to be expected, it is interesting to see that for fan-ins of more than~$16$, the traditional variant performs better, while for lower fan-ins, the shortcut variant is superior.
This effect can be explained by comparing the size of the virtual memory area that is accessed in both variants for $k$~slots referencing $m \leq k$ leaves.
Independent of the fan-in, the accesses on the shortcut variant always operate on a virtual memory area of size $k$~pages. In contrast to that, the traditional variant operates on a virtual memory area of only $k \cdot 8\text{B}$ for the inner node plus $m$ virtual pages for the leaf nodes.
Therefore, for higher fan-ins, the overhead of operating on a larger virtual memory area (causing more TLB misses and more expensive page table accesses) overshadows the benefit of eliminating indirections. 


Consequently, shortcuts should be primarily used under low fan-ins. Further, this implies that shortcut nodes should not entirely replace the traditional variant, but should be maintained \textit{alongside} with it.  
In Section~\ref{sec:eh}, we follow this by maintaining both the traditional and the shortcut variant and by guiding accesses through the best access path based on the current fan-in.  

\vspace*{-0.2cm}
\subsection{Factor in TLB-Shootdowns!}
\label{ssec:beware3}
\vspace*{-0.1cm}

When it comes to (re-)mapping, another relevant topic are so called TLB-shootdowns, which occur in the presence of multi-threaded applications. If a thread is performing \smalltt{mmap()} calls, i.e., to create a shortcut, then all existing outdated TLB entries must be invalidated to ensure correctness. This holds for both the thread-local TLB as well as the TLBs of all other running cores. 
Unfortunately, in contrast to data caches, TLBs do not implement a coherency mechanism in hardware. Therefore, the OS must issue so-called inter-processor interrupts (IPIs) to clear the outdated entries from all thread-remote TLBs, which is rather expensive~\cite{dontusemmap}.

\begin{figure}[h!]
\vspace*{-0.3cm}
\includegraphics[width=.87\linewidth, trim={0 0 0.6cm 0}, clip]{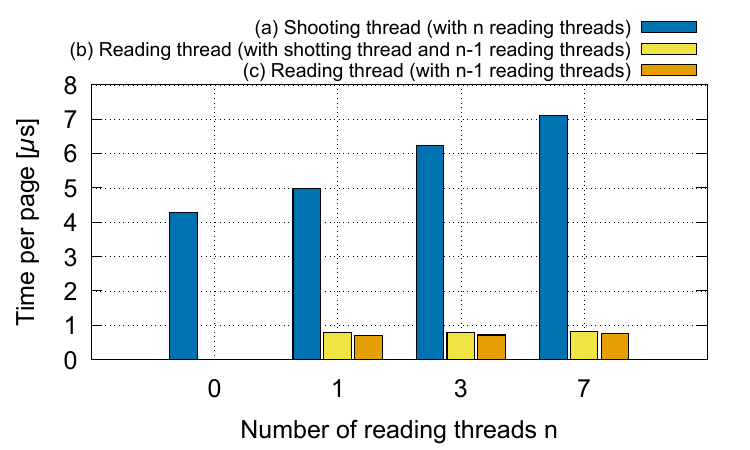}
\vspace*{-0.3cm}
\caption{Effect of TLB shootdowns.}
\label{fig:tlb_shootdowns}
\vspace*{-0.3cm}
\end{figure}

To understand the impact of TLB-shootdowns, we create a ``shooting'' thread that performs $2^{19}$ populated \smalltt{mmap()} calls on an already mapped memory region of size~$8$GB to remap $2^{19}$~randomly selected pages, causing a large number of TLB-shootdowns. While this is happening, a varying number of threads running on other cores are sequentially reading the memory region repeatedly until the shooting thread completed its task.
In Figure~\ref{fig:tlb_shootdowns}, we (a)~report the time it takes the shooting thread to remap one page as well as (b)~the time it takes one of the reading threads to read one page.
Additionally, as we count the total number of pages being read during the measurement of (a) and (b), we let the reading threads read the same amount of pages again but this time without a shooting thread intervening. Again, we (c)~report the time it takes one of the reading threads to read one page.

In the results we can first of all see that the cost of remapping a page (bar~(a)) increases significantly with the number of concurrently reading threads: With seven concurrently reading threads, the very same page remap costs $1.67\times$ more than without any reading threads operating. From the perspective of a reading thread (bar~(b)), this is not the case, i.e., the runtime of reading a page remains independent from the total number of reading threads. These observations are interesting, as they indicate that TLB shootdowns do not affect the threads being targeted, but actually slow down the shooting thread. 
When comparing  bar~(b) and bar~(c), we also see that only little overhead is caused by the shooting thread.

In summary, TLB-shootdowns slow down the thread that is performing the remapping. Therefore, as also identified in Section~\ref{ssec:beware1}, the cost of creating and updating the shortcut should be hidden.

\vspace*{-0.2cm}
\section{Cutting Short Extendible Hashing}
\label{sec:eh}
\vspace*{-0.1cm}

Finally, let us incorporate shortcuts into an actual database index, namely extendible hashing, which distributes the resizing cost over the sequence of insertions. It is a well-suited candidate for such an enhancement: (a)~It maintains a wide inner node (called \textit{directory}). (b)~It indexes larger fixed size leaves (called \textit{buckets}). (c)~Its disadvantage over a single hash table is the directory indirection.


\begin{figure}[h!]
\vspace*{-0.3cm}
\begin{subfigure}[b]{.40\linewidth}
\includegraphics[width=\linewidth]{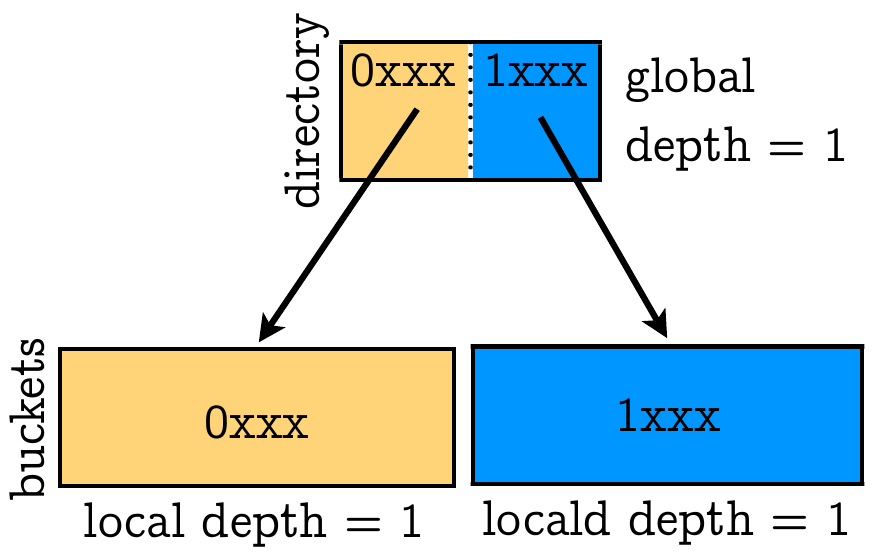}
\caption{Initial situation.}
\label{figs:eh_step1}
\end{subfigure}
\begin{subfigure}[b]{.58\linewidth}
\includegraphics[width=\linewidth]{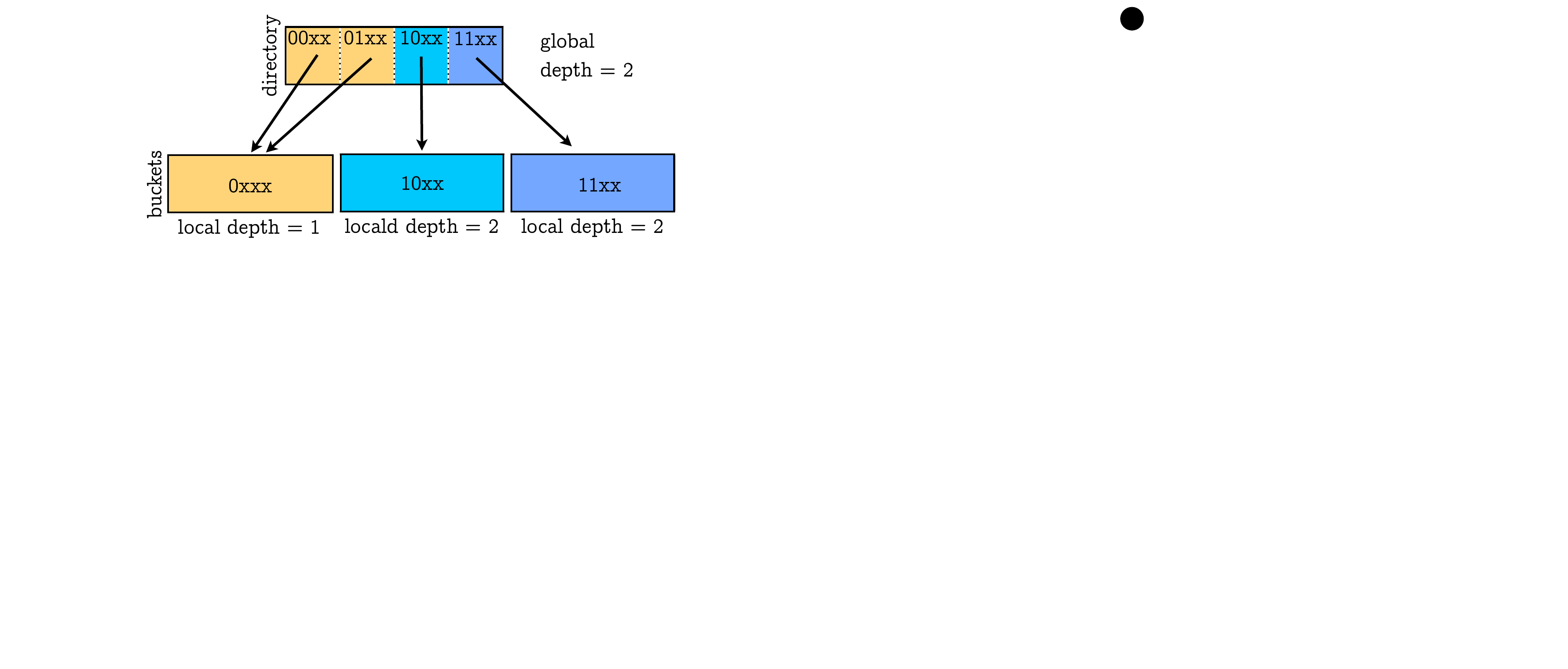}
\caption{After split of bucket \smalltt{1xxx}.}
\label{figs:eh_step2}
\end{subfigure}
\vspace*{-0.3cm}
\caption{The core principle of extendible hashing.}
\label{figs:extendible_hashing}
\vspace{-0.4cm}
\end{figure}

Figure~\ref{figs:extendible_hashing} shows its basic principle. Assuming that a hash consists of four bits, in Figure~\ref{figs:eh_step1}, we start with two buckets where the left one contains only entries whose hashes have the form~\smalltt{0xxx}, whereas the right one contains only entries whose hashes look like~\smalltt{1xxx}.
Thus, the \textit{local depth} of each bucket is~$1$, as only one bit of the hash determines to which bucket each entry belongs. Correspondingly, the  \textit{global depth} of the directory with two slots is~$1$. 
Let us now assume that the right bucket overflows. To handle the overflow, the bucket must be split and the entries must be moved and rehashed to two new buckets storing entries whose hashes look like \smalltt{10xx} and \smalltt{11xx}. Figure~\ref{figs:eh_step2} shows the result of this split. Consequently, the local depth of these two new buckets is now~$2$, as two bits of the hash are considered. The directory must be doubled too in order to index the new buckets, which increases the global depth to~$2$. The adaptiveness of the approach becomes visible at the untouched bucket \smalltt{0xxx}. Due to the doubling of the directory, this bucket is now referenced by the two directory slots, as the global depth is larger than the local depth of bucket~\smalltt{0xxx}. Also note that a split of bucket ~\smalltt{0xxx} would now \textit{not} double the directory, as enough slots are available to reference the two resulting buckets. 

\vspace*{-0.2cm}
\subsection{Architecture and Implementation}
\label{ssec:architecture}
\vspace*{-0.1cm}

To eliminate the major weakness of extendible hashing, in the following, we will cut the directory short \textit{while} keeping its adaptiveness intact.
Based on the lessons learned in Section~\ref{sec:micro_benchmarks}, we design our \textbf{Shortcut-EH} as follows: To hide the cost of creation and maintenance (Section~\ref{ssec:beware1}) and to reduce the slowdown caused by TLB-Shootdowns (Section~\ref{ssec:beware3}), the shortcut directory cannot replace the traditional directory entirely, but must accompany it. 
While all directory-modifying operations are reflected synchronously by the traditional directory, the shortcut directory replays these operations asynchronously. When the shortcut directory is in sync with the traditional directory, it is considered for accesses as a faster alternative.  
As a side-effect of this design, we can freely switch between traditional and shortcut directory for routing accesses, where we base the decision on the current average fan-in (Section~\ref{ssec:beware2}).

We trigger and coordinate the asynchronous maintenance of the shortcut directory via a concurrent lock-free FIFO queue. This queue receives maintenance requests from the main thread as soon as modifications on the traditional directory happen. These modifications can be of two types: 
(a)~Splitting a bucket. In this case, two slots of the traditional directory are updated to index two new buckets after a split and two remappings must happen in the corresponding shortcut directory. To do so, the main thread will push two \textit{update requests} into the queue when a bucket reorganization happens, each containing the slot to update as well as file offset to map the slot to. 
(b)~Doubling the traditional directory. In this case, the existing shortcut directory is destroyed and a new shortcut directory is created from scratch. To trigger this, the main thread puts a \textit{create request} on the queue containing the number of slots of the new directory as well as a vector of file offsets to map the slots to. 
Note that before the main thread pushes a create requests into the queue, it pops all potentially pending update requests as they became outdated. 
A separate mapper thread constantly polls the concurrent queue at a fixed frequency for requests, where we empirically determined $25$ms to work well in practice. If a request is pending, it executes it to update respectively replace the current shortcut directory. The execution of the requests is always followed by a corresponding page table population to ensure that all page table entries exist before any access happen. 

As mentioned, we can use an existing shortcut directory only, if it is in sync with the traditional directory. To detect synchronicity, we maintain for both the traditional as well as the shortcut directory a \textit{version number}, where every modification to a directory increments the respective version. 
We update the version number of the shortcut directory only after the population of the page table has been performed to assure that no access on the shortcut directory suffers from an expensive page fault. 
Note that even if the shortcut directory is in sync, we might not use it for every access due to the danger of TLB-thrashing. To make a decision, we keep the the current average fan-in of the directory and base our decision on that: 
If the average fan-in is~$\leq 8$, we route the access through the shortcut, otherwise, we use the traditional directory.

\vspace*{-0.2cm}
\subsection{Experimental Evaluation}
\label{ssec:experiments}
\vspace*{-0.1cm}

Let us now see how \textbf{Shortcut-EH} compares against the following baselines:
Hash Table~\textbf{(HT)}~is a single open-addressing/linear probing hash table with $n$~slots. If the load factor exceeds a specified theshold, a new hash table of size~$2n$ is allocated and all entries are rehashed to the new table.
Hash Table Incremental~\textbf{(HTI)}, as implemented by the popular key-value store Redis~\cite{redis_incremental_rehash}, ~resembles \textbf{HT} in all aspects except that instead of rehashing all entries over to the new hash table in one go, only a batch of~$b \leq n$ entries is moved. Subsequent accesses then also move~$b$ entries until everything is migrated. As long as both tables co-exist, lookups have to potentially inspect both tables for keys (starting with the one containing more entries).
Chained Hashing~\textbf{(CH)}~uses a fixed size hash table, where a slot either contains an entry or a pointer to a fixed-size bucket. When a bucket overflows, it creates a new bucket, links to it, and inserts the entry there. Buckets are searched linearly.
\textbf{Extendible Hashing~(EH)} resembles a classical extendible hashing scheme, which uses a pointer-based directory. The directory is indexed using the most significant bits of the key. Within each bucket, we use again open addressing/linear probing. 
\textbf{Shortcut Extendible Hashing~(\veh{})} resembles our shortcut-enhanced method described in Section~\ref{ssec:architecture}.
To ensure comparability, all methods utilize the same lightweight multiplicative hash function internally and a bucket size of $4$KB.

\begin{figure}[h!]
  \centering
  \begin{subfigure}[b]{\linewidth}
  \hspace*{0.2cm}
    \includegraphics[width=\linewidth, trim={0 0 0 0}, clip]{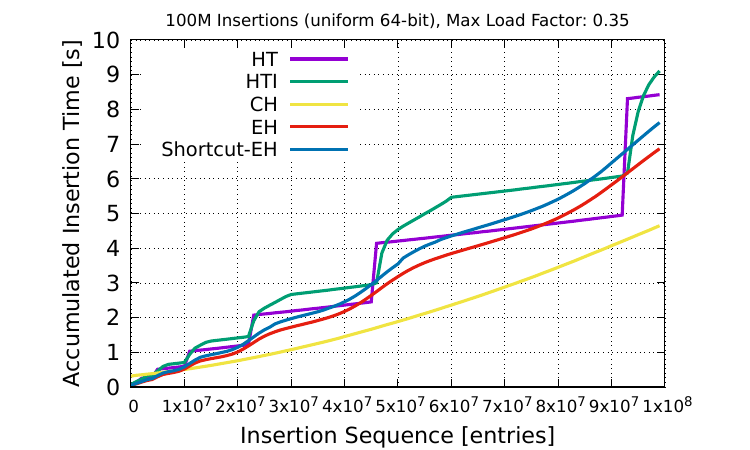}
    \vspace*{-0.3cm}
    \caption{Insertions.}
    \label{fig:insertions}
  \end{subfigure}
  
  \begin{subfigure}[b]{.92\linewidth}
    \includegraphics[width=\linewidth, trim={0 0.3cm 0 0}, clip]{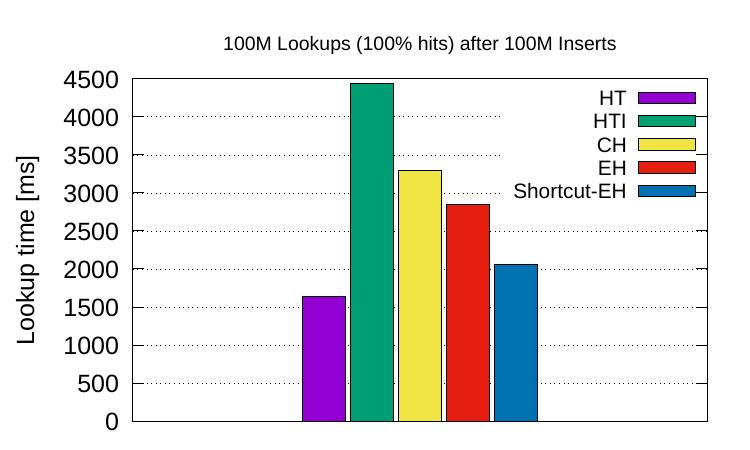}
     \vspace*{-0.3cm}
    \caption{Lookups.}
    \vspace*{-0.2cm}
    \label{fig:lookups}
  \end{subfigure}
  \caption{Insertion and lookup performance of hash tables.}
  \label{fig:uniform}
  \vspace*{-0.3cm}
\end{figure}

In Figure~\ref{fig:insertions}, we start by inserting $100$M entries into each index and report the accumulated insertion time. \textbf{HT}, \textbf{HTI}, \textbf{EH}, and \textbf{Shortcut-EH} all start with an effective space of only $4$KB and resize at a load factor of $35\%$.
As \textbf{CH} does not adjust its hash table size, it is allowed to start with a hash table size of $1$GB and links buckets of size~$128$B.
From the results we can see that both \textbf{EH} and \textbf{Shortcut-EH} indeed gracefully distribute the insertion cost over the sequence. This is clearly not the case for \textbf{HT} which shows a staircase shape due to the occasional doubling of the entire hash table. \textbf{HTI} reduces this problem by prolonging the rehashing step at the cost of keeping two hash tables side by side for a longer period of time. Unsurprisingly, \textbf{CH} shows the best insertion time, as it does not perform any rehashing at all. Most importantly for us, we can see that the overhead caused by the maintenance of the shortcut in \textbf{Shortcut-EH} is only around $8\%$ over~\textbf{EH}. 

Next, in Figure~\ref{fig:lookups}, we perform $100$M random lookups (only hits) on the previously filled indexes. Note that for \textbf{Shortcut-EH}, the shortcut is in sync with the traditional directory and hence used for all lookups.
Consequently, \textbf{Shortcut-EH} performs significantly better than \textbf{EH}, reducing its major limitation. Also, while \textbf{HT} offers the fastest lookups, it is closely followed by \textbf{Shortcut-EH}. The small overhead is caused by having to compute two hashes (directory slot and bucket slot) instead of only one hash per lookup as well as by having to test for whether the directories are in sync. 
\textbf{CH} and in particular \textbf{HTI} pay a price for having to traverse bucket lists respectively accessing two hash tables to locate the entry.   

\begin{figure}[h!]
\centering
\vspace*{-0.3cm}
\includegraphics[width=.95\linewidth, trim={0 0 0 0}, clip]{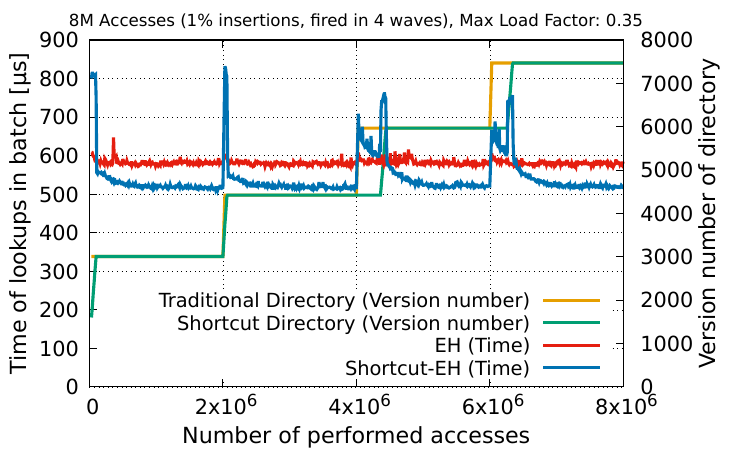}
\vspace*{-0.4cm}
\caption{Synchronization under a mixed workload.}
\label{fig:hybrid}
\vspace*{-0.4cm}
\end{figure}

Finally, in Figure~\ref{fig:hybrid}, let us inspect the behavior of \textbf{Shortcut-EH} in comparison with \textbf{EH} in more detail under a mixed workload containing both insertions and lookups. In particular, we are inserted in the synchronization of the shortcut with its traditional counterpart and its effect on the lookup performance.
To do so, upfront, we first bulk-load both indexes with $92$M entries. Then, we fire four waves of $2$M accesses each, where the first $1\%$~accesses are insertions and the remaining $99\%$~accesses are lookups, resembling a read-heavy workload. We plot the lookup time for every $10.000$~accesses. Along, we show the current version number of both directories to see when and for how long they are out of sync.
We can see that the insertions happening at the beginning of each wave trigger bucket splits and cause the shortcut directory to go out of sync. This penalizes the lookup time of \textbf{Shortcut-EH}, as the lookups must be answered by the traditional directory concurrently to the synchronization. However, shortly after the insertion burst ends, the shortcut directory catches up and the lookup time of \textbf{Shortcut-EH} clearly falls below the one of \textbf{EH} again.

\vspace*{-0.2cm}
\section{Conclusion}
\label{sec:conclusion}
\vspace*{-0.1cm}

In this paper we introduced virtual memory shortcuts to reduce the number of indirections within radix-style index structures. We presented the technique and experimentally analyzed its behavior during creation, varying fan-ins, and concurrency. Based on these insights, we exemplarily integrated shortcuts into extendible hashing and showed that the technique eliminates the major downside of the hashing scheme, namely the overhead caused by directory, narrowing the gap to single table hashing. 

\bibliographystyle{ACM-Reference-Format}
\bibliography{virtual_indexing}


\begin{thebibliography}{10}


\ifx \showCODEN    \undefined \def \showCODEN     #1{\unskip}     \fi
\ifx \showDOI      \undefined \def \showDOI       #1{#1}\fi
\ifx \showISBNx    \undefined \def \showISBNx     #1{\unskip}     \fi
\ifx \showISBNxiii \undefined \def \showISBNxiii  #1{\unskip}     \fi
\ifx \showISSN     \undefined \def \showISSN      #1{\unskip}     \fi
\ifx \showLCCN     \undefined \def \showLCCN      #1{\unskip}     \fi
\ifx \shownote     \undefined \def \shownote      #1{#1}          \fi
\ifx \showarticletitle \undefined \def \showarticletitle #1{#1}   \fi
\ifx \showURL      \undefined \def \showURL       {\relax}        \fi
\providecommand\bibfield[2]{#2}
\providecommand\bibinfo[2]{#2}
\providecommand\natexlab[1]{#1}
\providecommand\showeprint[2][]{arXiv:#2}

\bibitem[red(2020)]%
        {redis_incremental_rehash}
 \bibinfo{year}{2020}\natexlab{}.
\newblock
\newblock
\urldef\tempurl%
\url{https://codeburst.io/a-closer-look-at-redis-dictionary-implementation-internals-3fd815aae535}
\showURL{%
\tempurl}


\bibitem[Crotty et~al\mbox{.}(2022)]%
        {dontusemmap}
\bibfield{author}{\bibinfo{person}{Andrew Crotty}, \bibinfo{person}{Viktor
  Leis}, {and} \bibinfo{person}{Andrew Pavlo}.}
  \bibinfo{year}{2022}\natexlab{}.
\newblock \showarticletitle{Are You Sure You Want to Use {MMAP} in Your
  Database Management System?}. In \bibinfo{booktitle}{\emph{{CIDR} 2022,
  Chaminade, CA, USA, January 9-12, 2022}}.
\newblock


\bibitem[Fagin et~al\mbox{.}(1979)]%
        {eh}
\bibfield{author}{\bibinfo{person}{Ronald Fagin} {et~al\mbox{.}}}
  \bibinfo{year}{1979}\natexlab{}.
\newblock \showarticletitle{Extendible Hashing - {A} Fast Access Method for
  Dynamic Files}.
\newblock \bibinfo{journal}{\emph{{ACM} Trans. Database Syst.}}
  \bibinfo{volume}{4}, \bibinfo{number}{3} (\bibinfo{year}{1979}),
  \bibinfo{pages}{315--344}.
\newblock


\bibitem[Haffner et~al\mbox{.}(2018)]%
        {rewired_cracking}
\bibfield{author}{\bibinfo{person}{Immanuel Haffner},
  \bibinfo{person}{Felix~Martin Schuhknecht}, {and} \bibinfo{person}{Jens
  Dittrich}.} \bibinfo{year}{2018}\natexlab{}.
\newblock \showarticletitle{An analysis and comparison of database cracking
  kernels}. In \bibinfo{booktitle}{\emph{DaMoN, Houston, TX, USA, June 11,
  2018}}. \bibinfo{publisher}{{ACM}}, \bibinfo{pages}{10:1--10:10}.
\newblock


\bibitem[Leo and Boncz(2019)]%
        {pma}
\bibfield{author}{\bibinfo{person}{Dean~De Leo} {and} \bibinfo{person}{Peter~A.
  Boncz}.} \bibinfo{year}{2019}\natexlab{}.
\newblock \showarticletitle{Packed Memory Arrays - Rewired}. In
  \bibinfo{booktitle}{\emph{{ICDE} 2019, Macao, China, April 8-11, 2019}}.
  \bibinfo{publisher}{{IEEE}}, \bibinfo{pages}{830--841}.
\newblock


\bibitem[Schuhknecht and Henneberg(2023a)]%
        {storageviews2}
\bibfield{author}{\bibinfo{person}{Felix Schuhknecht} {and}
  \bibinfo{person}{Justus Henneberg}.} \bibinfo{year}{2023}\natexlab{a}.
\newblock \showarticletitle{Accelerating Main-Memory Table Scans with Partial
  Virtual Views}. In \bibinfo{booktitle}{\emph{{DaMoN} 2023, Seattle, WA, USA,
  June 18-23, 2023}}. \bibinfo{publisher}{{ACM}}, \bibinfo{pages}{89--93}.
\newblock


\bibitem[Schuhknecht et~al\mbox{.}(2016)]%
        {rewiring}
\bibfield{author}{\bibinfo{person}{Felix~Martin Schuhknecht},
  \bibinfo{person}{Jens Dittrich}, {and} \bibinfo{person}{Ankur Sharma}.}
  \bibinfo{year}{2016}\natexlab{}.
\newblock \showarticletitle{{RUMA} has it: Rewired User-space Memory Access is
  Possible!}
\newblock \bibinfo{journal}{\emph{Proc. {VLDB} Endow.}} \bibinfo{volume}{9},
  \bibinfo{number}{10} (\bibinfo{year}{2016}), \bibinfo{pages}{768--779}.
\newblock


\bibitem[Schuhknecht and Henneberg(2023b)]%
        {storageviews3}
\bibfield{author}{\bibinfo{person}{Felix~Martin Schuhknecht} {and}
  \bibinfo{person}{Justus Henneberg}.} \bibinfo{year}{2023}\natexlab{b}.
\newblock \showarticletitle{Towards Adaptive Storage Views in Virtual Memory}.
  In \bibinfo{booktitle}{\emph{{CIDR} 2023, Amsterdam, The Netherlands, January
  8-11, 2023}}. \bibinfo{publisher}{www.cidrdb.org}.
\newblock


\bibitem[Schuhknecht et~al\mbox{.}(2021)]%
        {anyolap}
\bibfield{author}{\bibinfo{person}{Felix~Martin Schuhknecht},
  \bibinfo{person}{Aaron Priesterroth}, \bibinfo{person}{Justus Henneberg},
  {and} \bibinfo{person}{Reza Salkhordeh}.} \bibinfo{year}{2021}\natexlab{}.
\newblock \showarticletitle{AnyOLAP: Analytical Processing of Arbitrary
  Data-Intensive Applications without {ETL}}.
\newblock \bibinfo{journal}{\emph{Proc. {VLDB} Endow.}} \bibinfo{volume}{14},
  \bibinfo{number}{12} (\bibinfo{year}{2021}), \bibinfo{pages}{2823--2826}.
\newblock


\bibitem[Tamminen(1981)]%
        {multi_level_eh}
\bibfield{author}{\bibinfo{person}{Markku Tamminen}.}
  \bibinfo{year}{1981}\natexlab{}.
\newblock \showarticletitle{Order Preserving Extendible Hashing and Bucket
  Tries}.
\newblock \bibinfo{journal}{\emph{{BIT}}} \bibinfo{volume}{21},
  \bibinfo{number}{4} (\bibinfo{year}{1981}), \bibinfo{pages}{419--435}.
\newblock


\end{thebibliography}

\end{document}